\documentclass[a4paper,11pt]{article}
\usepackage{pos}

\usepackage{braket}
\usepackage{bbm}

\title{Numerical studies on the finite-temperature CP restoration in 4D SU(N) gauge theory at $\theta=\pi$}

\ShortTitle{Finite-temperature CP restoration in 4D SU(N) gauge theory}

\author*[a]{Akira Matsumoto}
\author[b]{Kohta Hatakeyama}
\author[c]{Mitsuaki Hirasawa}
\author[a,d]{Masazumi Honda}
\author[b,e]{Jun Nishimura}
\author[f]{Atis Yosprakob}

\affiliation[a]{RIKEN iTHEMS,\\ 2-1 Hirosawa, Wako, Saitama 351-0198, Japan}

\affiliation[b]{KEK Theory Center, High Energy Accelerator Research Organization,\\ 1-1 Oho, Tsukuba, Ibaraki 305-0801, Japan}

\affiliation[c]{Sezione di Milano Bicocca, Istituto Nazionale di Fisica Nucleare,\\ Piazza della Scienza, 3, I-20126 Milano, Italy}

\affiliation[d]{Center for Gravitational Physics and Quantum Information, Yukawa Institute for Theoretical Physics,\\ Kyoto University, Sakyo-ku, Kyoto 606-8502, Japan}

\affiliation[e]{The Graduate University for Advanced Studies, SOKENDAI,\\ 1-1 Oho, Tsukuba, Ibaraki 305-0801, Japan}

\affiliation[f]{Department of Physics, Niigata University, Niigata 950-2181, Japan.}

\emailAdd{akira.matsumoto@riken.jp}
\emailAdd{khat@post.kek.jp}
\emailAdd{mitsuaki.hirasawa@mib.infn.it}
\emailAdd{masazumi318@gmail.com}
\emailAdd{jnishi@post.kek.jp}
\emailAdd{ayosp@phys.sc.niigata-u.ac.jp}

\abstract{Recent studies on the 't Hooft anomaly matching condition have suggested
a nontrivial phase structure in 4D SU($N$) gauge theory at $\theta=\pi$.
In the large-$N$ limit, it has been found that CP symmetry at $\theta=\pi$ is broken
in the confined phase, while it restores in the deconfined phase,
which is indeed one of the possible scenarios.
However, at small $N$, one may find other situations that are consistent
with the consequence of the anomaly matching condition.
Here we investigate this issue for $N=2$ by direct lattice calculations.
The crucial point to note is that the CP restoration can be probed
by the sudden change of the tail of the topological charge distribution at $\theta=0$,
which can be seen by simulating the theory at imaginary $\theta$ without the sign problem.
Our results suggest that the CP restoration at $\theta=\pi$ occurs at temperature 
higher than the deconfining temperature unlike the situation in the large-$N$ limit.}

\FullConference{%
  The 39th International Symposium on Lattice Field Theory (Lattice2022),\\
  8-13 August, 2022 \\
  Bonn, Germany 
}

\begin{document}

\begin{flushright}
KEK-TH-2488, RIKEN-iTHEMS-Report-23
\end{flushright}

\maketitle


\section{Introduction}

The non-perturbative effect of the topological theta term in quantum
field theories has been studied as a long-standing problem. Recently,
the phase structure of 4D pure Yang-Mills (YM) theory with a theta term
has attracted a lot of attention. There was a novel progress on application
of \textquoteright t Hooft anomaly matching \cite{Gaiotto:2017yup,Kitano:2017jng},
which suggests that the 4D SU($N$) pure YM theory
cannot have a unique trivial vacuum at $\theta=\pi$. Indeed, this
statement is consistent with the known phase diagram for large $N$,
where the CP symmetry at $\theta=\pi$ is spontaneously broken in
the confined phase. On the other hand, the phase diagram for small
$N$, in particular $N=2$, is not determined yet. It is possible
that the SU(2) YM theory has a qualitatively different phase structure.

Thus, it is an interesting challenge to investigate the phase structure
by a first-principle method. Since the effect of the theta term is
genuinely non-perturbative, the SU($N$) YM theory with a theta term
should be analyzed by non-perturbative methods. However, usual Monte
Carlo simulations of the lattice gauge theory including the theta
term is difficult due to the sign problem.

We propose a new method to probe the critical behavior at $\theta=\pi$
based on the topological charge distribution $\rho(q)$ at $\theta=0$.
The crucial point of the method is that the expectation value $\langle Q\rangle_{\theta}$
of topological charge for any $\theta$ is completely determined by
the distribution $\rho(q)$. Thanks to this property, we can investigate
the behavior of $\langle Q\rangle_{\theta}$ indirectly by the information
at $\theta=0$. In order to see the temperature dependence of the
distribution $\rho(q)$ clearly, we introduce the imaginary $\theta$
parameter, which can enhance the tail structure of $\rho(q)$.

\section{Identifying the CP restoration}

The application of \textquoteright t Hooft anomaly matching condition \cite{Gaiotto:2017yup,Kitano:2017jng}
to the 4D pure YM theory suggests that the phase structure at $\theta=\pi$
should be nontrivial. There are a lot of possible phase structures
which agree with this condition. Here we consider two kinds of phase
transition. One is the deconfinement transition at $T=T_{\mathrm{dec}}(\theta=\pi)$,
which corresponds to breaking of $Z_{N}$ center symmetry. Note that
the deconfining temperature $T_{\mathrm{dec}}(\theta)$ depends on $\theta$
in general. The other transition is the restoration of CP symmetry
at $T=T_{\mathrm{CP}}$, which is broken at low temperature. For large
$N$, these two transitions occur at the same temperature $T_{\mathrm{CP}}=T_{\mathrm{dec}}(\pi)$.
Namely, the CP symmetry is recovered simultaneously with the deconfinement
transition. In this case, either the $Z_{N}$ center symmetry or the
CP symmetry is broken at any temperature. Thus, it is consistent with
the anomaly matching condition. 

It is interesting to explore whether the theory with $N=2$ has a similar phase diagram. In fact, the numerical study of 4D SU(2) YM theory by the
subvolume method \cite{Kitano:2020mfk,Kitano:2021jho} shows an indication
of the CP broken phase at low temperature. It is also confirmed that
the instanton gas phase, which is CP symmetric, appears at high temperature.
Thus, the restoration of CP symmetry is expected to occur also for
$N=2$. However, the relation between the two critical temperatures
$T_{\mathrm{CP}}$ and $T_{\mathrm{dec}}(\pi)$ can be different.
The anomaly matching condition for these two temperatures requires that $T_{\mathrm{CP}}\geq T_{\mathrm{dec}}(\pi)$.
The reason is that the CP symmetry can be broken not only in the confined
phase but also in the deconfined phase. The overlap of the CP broken
phase and the $Z_{2}$ broken phase is allowed. There is a related
study of these two critical temperatures by using the super YM theory
\cite{Chen:2020syd}. Interestingly, the result shows $T_{\mathrm{CP}}>T_{\mathrm{dec}}(\pi)$
only for $N=2$, but $T_{\mathrm{CP}}=T_{\mathrm{dec}}(\pi)$ for
$N\geq3$.

It is worth trying to numerically investigate the CP restoration temperature
$T_{\mathrm{CP}}$ for the 4D SU(2) YM theory and compare it with
the deconfining temperature $T_{\mathrm{dec}}(\pi)$. However, the usual
Monte Carlo simulation at $\theta=\pi$ suffers from the sign problem
since the theta term is purely imaginary. In this section we introduce
a new method to determine $T_{\mathrm{CP}}$ without direct simulation
at $\theta=\pi$. 

First, we explain the property of topological charge $Q$. Since the
topological charge is a CP odd operator, its expectation value $\langle Q\rangle_{\theta}$
can be an order parameter of CP symmetry. 
\begin{equation}
\langle Q\rangle_{\theta}=-i\frac{\partial}{\partial\theta}\log Z_{\theta}
\end{equation}
If CP symmetry at $\theta=\pi$ is spontaneously broken, $\langle Q\rangle_{\theta}$
should be discontinuous there:
\begin{equation}
\Delta Q=\Big|\langle Q\rangle_{\theta=\pi-\epsilon}-\langle Q\rangle_{\theta=\pi+\epsilon}\Big|\begin{cases}
>0 & :\mathrm{CP\,broken},\\
=0 & :\mathrm{CP\,restored}.
\end{cases}
\end{equation}
Thus, the CP restoration temperature $T_{\mathrm{CP}}$ can be regarded
as a temperature at which $\Delta Q$ vanishes. To determine $T_{\mathrm{CP}}$,
we need to investigate the temperature dependence of $\Delta Q$.
However, it is difficult to directly evaluate $\Delta Q$ due to
the sign problem. But we can also study it from another direction. Let us note that the partition
function $Z_{\theta}$ and the topological charge distribution $\rho(q)$
at $\theta=0$ are related via 
\begin{equation}
\rho(q)=\frac{1}{Z_{0}}\int dA\,\delta(q-Q)e^{-S_{g}}=\frac{1}{Z_{0}}\int\frac{d\theta}{2\pi}\,e^{-i\theta q}Z_{\theta}.
\end{equation}
Since the distribution $\rho(q)$ is a Fourier transform of the partition
function $Z_{\theta}$, 
\begin{equation}
Z_{\theta}=\int dA\,e^{-S_{g}+i\theta Q}=Z_{0}\int dq\,e^{i\theta q}\rho(q),
\end{equation}
we find that the expectation value $\langle Q\rangle_{\theta}$ at
any $\theta$ is completely determined by $\rho(q)$ as 
\begin{equation}
\langle Q\rangle_{\theta}=-i\frac{\partial}{\partial\theta}\log\int dq\,e^{i\theta q}\rho(q)=\frac{\int dq\,qe^{i\theta q}\rho(q)}{\int dq\,e^{i\theta q}\rho(q)}.
\end{equation}
This is nothing but the reweighting formula by using the information
at $\theta=0$. Thus, to calculate $\langle Q\rangle_{\theta}$ around
$\theta\sim\pi$, we need exponentially large amount of statistics.
However, our goal is not to determine the complete $\theta$ dependence
of $\langle Q\rangle_{\theta}$ but to probe the critical temperature
$T_{\mathrm{CP}}$. In fact, we do not need the complete information
of $\rho(q)$ in that case. It is enough to determine whether $\Delta Q$
is zero or not from $\rho(q)$. 

In this study, we propose to use the expectation value $\left\langle Q\right\rangle _{\tilde{\theta}}$
of the topological charge at an imaginary theta $\theta=i\tilde{\theta}$
($\tilde{\theta}\in\mathbb{R}$) to probe the critical behavior;
\begin{equation}
\left\langle Q\right\rangle _{\tilde{\theta}}=\frac{1}{Z_{\tilde{\theta}}}\int dA\,Qe^{-S_{g}-\tilde{\theta}Q}=\frac{\int dq\,qe^{-\tilde{\theta}q}\rho(q)}{\int dq\,e^{-\tilde{\theta}q}\rho(q)}.\label{eq:Q_imag_theta}
\end{equation}
In practice, we normalize it by the topological susceptibility 
\begin{equation}
\chi_{0}=\frac{1}{V}\langle Q^{2}\rangle_{\theta=0}
\end{equation}
at $\theta=0$ and the volume $V$, so that we just measure the ratio
of two independent observables. 
\begin{equation}
\frac{\left\langle Q\right\rangle _{\tilde{\theta}}}{\chi_{0}V}=\frac{\left\langle Q\right\rangle _{\tilde{\theta}}}{\langle Q^{2}\rangle_{0}}
\end{equation}
Let us discuss the behavior of this observable in some well known models.
The first example is the instanton gas model, for which the free energy
is obtained as 
\begin{equation}
F_{\theta}=-\log Z_{\theta}=\chi_{0}V(1-\cos\theta).
\end{equation}
The $\theta$-dependence of $\left\langle Q\right\rangle _{\theta}/\chi_{0}V$
for real $\theta$ is given by the sine function 
\begin{equation}
\frac{\left\langle Q\right\rangle _{\theta}}{\chi_{0}V}=i\sin\theta,
\end{equation}
which indicates that CP symmetry at $\theta=\pi$ is not broken.
Correspondingly, the imaginary-$\theta$ dependence turns out to be
the hyperbolic sine function. 

\begin{equation}
\frac{\left\langle Q\right\rangle _{\tilde{\theta}}}{\chi_{0}V}=-\sinh\tilde{\theta}
\end{equation}
The second example is the Gaussian model 
\begin{equation}
F_{\theta}=\frac{1}{2}\chi_{0}V\min_{n}(\theta-2\pi n)^{2},
\end{equation}
which is known to be realized for large $N$ at low temperature.
The real-$\theta$ dependence of $\left\langle Q\right\rangle _{\theta}/\chi_{0}V$
is given by 
\begin{equation}
\frac{\left\langle Q\right\rangle _{\theta}}{\chi_{0}V}=i(\theta \mod 2\pi)
\end{equation}
for $V\gg1$, which indicates that the CP is broken. For imaginary $\theta$,
we find the linear behavior. 
\begin{equation}
\frac{\left\langle Q\right\rangle _{\tilde{\theta}}}{\chi_{0}V}=-\tilde{\theta}
\end{equation}
We can see the clear difference between the behaviors of $\left\langle Q\right\rangle _{\tilde{\theta}}/\chi_{0}V$
for these two models. It behaves as $-\sinh\tilde{\theta}$ for the
instanton gas model (CP restored), while it behaves as $-\tilde{\theta}$
for the Gaussian model (CP broken). Although the 4D SU(2) YM theory
will not be as simple as these models, this observable is still useful
to investigate the CP restoration. In fact, the expectation value
$\left\langle Q\right\rangle _{\tilde{\theta}}$ for imaginary $\theta$
is sensitive to the tail of the distribution of $\rho(q)$. The imaginary
theta term enhances the contribution of large-$q$ sectors because
of the factor $e^{-\tilde{\theta}q}$ in the integrand of (\ref{eq:Q_imag_theta}).
Note that, for these two examples, the tail of $\rho(q)$ behaves
for $q\gg1$ as follows:
\begin{equation}
\rho(q)\sim\begin{cases}
\exp\left(-q\log\frac{2q}{\chi_{0}V}\right) & :\mathrm{instanton\,gas},\\
\exp\left(-\frac{q^{2}}{2\chi_{0}V}\right) & :\mathrm{Gaussian}.
\end{cases}
\end{equation}

\section{4D SU(2) gauge theory with a theta term}

In this study, we focus on the SU(2) pure Yang-Mills theory on
the 4D Euclidean space. The action for the gauge field $A_{\mu}^{a}$
($a=1,2,3$) ($\mu=1,\dots,4$) is defined by 
\begin{equation}
S_{g}=\frac{1}{4g^{2}}\int d^{4}x\ F_{\mu\nu}^{a}F_{\mu\nu}^{a},
\end{equation}
where $g$ is the coupling constant and $F_{\mu\nu}^{a}$ is the field
strength. 
\begin{equation}
F_{\mu\nu}^{a}=\partial_{\mu}A_{\nu}^{a}-\partial_{\nu}A_{\mu}^{a}-\epsilon^{abc}A_{\mu}^{b}A_{\nu}^{c}
\end{equation}
The topological charge is given by 
\begin{equation}
Q=\frac{1}{64\pi^{2}}\int d^{4}x\ \epsilon_{\mu\nu\rho\sigma}F_{\mu\nu}^{a}F_{\rho\sigma}^{a},\label{eq:Q}
\end{equation}
which takes an integer value on the compact space. We introduce the
topological theta term $S_{\theta}=-i\theta Q$ with a parameter $\theta\in\mathbb{R}$,
so that the total action is $S=S_{g}+S_{\theta}$. Since the partition
function 
\begin{equation}
Z=\int\mathcal{D}A\ e^{-S_{g}+i\theta Q}
\end{equation}
is invariant under the shift $\theta\rightarrow\theta+2\pi$, the
theory has $2\pi$ periodicity with respect to $\theta$. 
Since the parameter $\theta$
flips its sign by the CP transformation $\theta\rightarrow-\theta$,
the theta term explicitly breaks the CP symmetry for $\theta\neq0$.
However, thanks to the $2\pi$ periodicity, the CP symmetry exists
also at $\theta=\pi$.

Next, we define the lattice action for the numerical study. The gauge
field is represented by the link variable $U_{n,\mu}\in\mathrm{SU}(2)$.
The index $n$ labels the lattice sites. The plaquette is given by
\begin{equation}
P_{n}^{\mu\nu}=U_{n,\mu}U_{n+\hat{\mu},\nu}U_{n+\hat{\nu},\mu}^{\dagger}U_{n,\nu}^{\dagger},
\end{equation}
where $\hat{\mu}$ represents the unit vector along the $\mu$-th
direction. Then we define the plaquette action with the lattice coupling
constant $\beta$. 
\begin{equation}
S_{g}=-\frac{\beta}{4}\sum_{n}\sum_{\mu\neq\nu}\mathrm{Tr}(P_{n}^{\mu\nu})
\end{equation}
Similarly, we can define the topological charge on the lattice by
the so-called \textquotedbl clover leaf\textquotedbl{} formula,
\cite{DiVecchia:1981aev} 
\begin{equation}
Q_{\textrm{cl}}=-\frac{1}{32\pi^{2}}\sum_{n}\frac{1}{2^{4}}\sum_{\mu,\nu,\rho,\sigma=\pm1}^{\pm4}\tilde{\epsilon}_{\mu\nu\rho\sigma}\mathrm{Tr}(P_{n}^{\mu\nu}P_{n}^{\rho\sigma}).\label{eq:Qcl}
\end{equation}
Here the orientation of the plaquette is extended to the negative
directions as well. The corresponding anti-symmetric tensor $\tilde{\epsilon}_{\mu\nu\rho\sigma}$
also has negative indices, so that 
\begin{equation}
1=\tilde{\epsilon}_{1234}=-\tilde{\epsilon}_{2134}=-\tilde{\epsilon}_{-1234}=\cdots.
\end{equation}
 It is known that the naively defined topological charge $Q_{\textrm{cl}}$
does not take an integer value on the lattice due to the discretization
effect. In order to recover the topological property of the gauge
field, we need to eliminate short-range fluctuations. In fact, there
are some smoothing techniques, such as the gradient flow, stout smearing
and so on. By using such a technique, we can define the smeared topological
charge so that it becomes close to an integer. In this study, we introduce the stout smearing
to the hybrid Monte Carlo simulation, which is discussed in section
\ref{sec:Stout-smearing}.

\section{Stout smearing for the HMC \label{sec:Stout-smearing}}

Since the CP symmetry at $\theta=\pi$ is related to the $2\pi$ periodicity
of $\theta$, the topological property of the theory is essential in the study of the phase structure. Thus, we use the stout smearing \cite{Morningstar:2003gk}
to define the topological charge. In the hybrid Monte Carlo simulation,
the drift force is used to update the configuration. If the action
has the theta term with the smeared topological charge, it also contributes
to the drift force. We can explicitly calculate the drift force from
the smeared topological charge by using stout smearing.  In this
section, we briefly review the stout smearing in the hybrid Monte
Carlo simulation. 

Stout smearing is an iterative procedure to obtain the smeared
link $\tilde{U}_{n,\mu}$ starting from the original link $U_{n,\mu}$.
We call the number of iterations $N_{\rho}$. 
\begin{equation}
U_{n,\mu}=U_{n,\mu}^{(0)}\rightarrow U_{n,\mu}^{(1)}\rightarrow\cdots\rightarrow U_{n,\mu}^{(N_{\rho})}=\tilde{U}_{n,\mu}.\label{eq:smearing_steps}
\end{equation}
In one (isotropic) smearing step from $k$ to $k+1$, the link variable
$U_{n,\mu}^{(k)}\in\mathrm{SU}(2)$ is mapped to $U_{n,\mu}^{(k+1)}\in\mathrm{SU}(2)$
defined by following formulae:
\begin{equation}
U_{n,\mu}^{(k+1)}=e^{iY_{n,\mu}}U_{n,\mu}^{(k)},
\end{equation}
\begin{equation}
iY_{n,\mu}=-\frac{\rho}{2}\mathrm{Tr}(J_{n,\mu}\tau^{a})\tau^{a}\label{eq:Y},
\end{equation}
\begin{equation}
J_{n,\mu}=U_{n,\mu}\Omega_{n,\mu}-\Omega_{n,\mu}^{\dagger}U_{n,\mu}^{\dagger},
\end{equation}
\begin{equation}
\Omega_{n,\mu}=\sum_{\sigma(\neq\mu)}\left(U_{n+\hat{\mu},\sigma}U_{n+\hat{\sigma},\mu}^{\dagger}U_{n,\sigma}^{\dagger}+U_{n+\hat{\mu}-\hat{\sigma},\sigma}^{\dagger}U_{n-\hat{\sigma},\mu}^{\dagger}U_{n-\hat{\sigma},\sigma}\right).\label{eq:OMG}
\end{equation}
Here $\tau^{a}$ are the SU(2) generators in fundamental representation.
The smearing step parameter $\rho>0$ should be chosen appropriately
depending on the system. 

In the hybrid Monte Carlo simulation,
we obtain the smeared link $\tilde{U}_{n,\mu}$
by this procedure, and then we use $\tilde{U}_{n,\mu}$ to calculate
the topological charge (\ref{eq:Qcl}) instead of the original link
$U_{n,\mu}$. The topological charge given by the stout smearing  
\begin{equation}
Q:=Q_{\textrm{cl}}(\tilde{U})\label{eq:def_Q}
\end{equation}
is used in the theta term $S_{\theta}=-i\theta Q$ as well as
in measuring the observable. In the step of molecular dynamics, we need
to calculate the drift force 
\begin{equation}
F_{n,\mu}=i\tau^{a}D_{n,\mu}^{a}S_{\theta}\label{eq:drift_theta}
\end{equation}
from the theta term. Although $S_{\theta}$ is a complicated function
of the original link variable $U_{n,\mu}$, it is possible to calculate
the drift force by reversing the smearing steps (\ref{eq:smearing_steps}).

\section{Result of the HMC}

\begin{figure}
\begin{centering}
\includegraphics[scale=0.65]{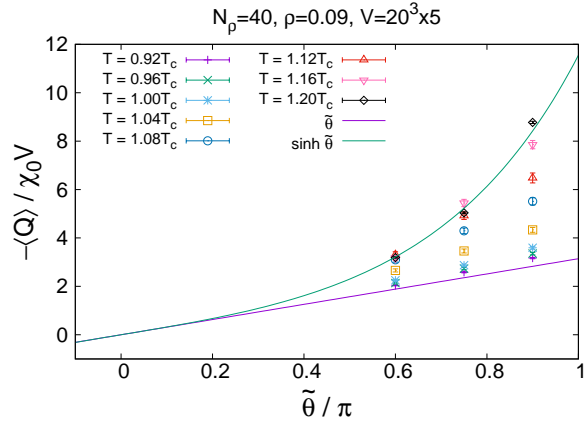}
\par\end{centering}
\caption{\label{fig:Q_theta_var_temp} The imaginary $\theta$ dependence of
$-\left\langle Q\right\rangle _{\tilde{\theta}}/\chi_{0}V$ for various
values of temperature in $0.92\le T/T_{{\rm dec}}(0)\le1.20$ with increments
of 0.04. The green solid curve is obtained by the instanton gas approximation,
which is valid at high temperature. The purple solid line is obtained
by the Gaussian model.}
\end{figure}

In this section, we show the result of the hybrid Monte Carlo simulation
with the imaginary theta term. For the stout smearing, we set $N_{\rho}=40$
and $\rho=0.09$ so that the topological charge is close to an
integer. In Fig.~\ref{fig:Q_theta_var_temp}, we plot $-\langle Q\rangle_{\tilde{\theta}}/\chi_{0}V$
against $\tilde{\theta}/\pi=\theta/i\pi$ for various values of temperature
in the range $0.9\leq T/T_{{\rm dec}}(\theta=0)\leq1.2$. We found
that, at high temperature, the data points are consistent with the
instanton gas approximation. On the other hand, the data points approach
the behavior of the Gaussian model at low temperature. It is convincing
that the SU(2) YM theory behaves as the instanton gas model at high temperature.
However, it does not necessarily coincides with the Gaussian model at
low temperature since the situation of $N=2$ can be different from
that of large $N$. Nevertheless, this observable is suitable
for probing the phase structure. 
Indeed, we can see that the behaviors of $\langle Q\rangle_{\tilde{\theta}}/\chi_{0}V$
change drastically slightly above the deconfining temperature $T_{{\rm dec}}(0)$
at $\theta=0$.

\begin{figure}
\begin{centering}
\includegraphics[scale=0.55]{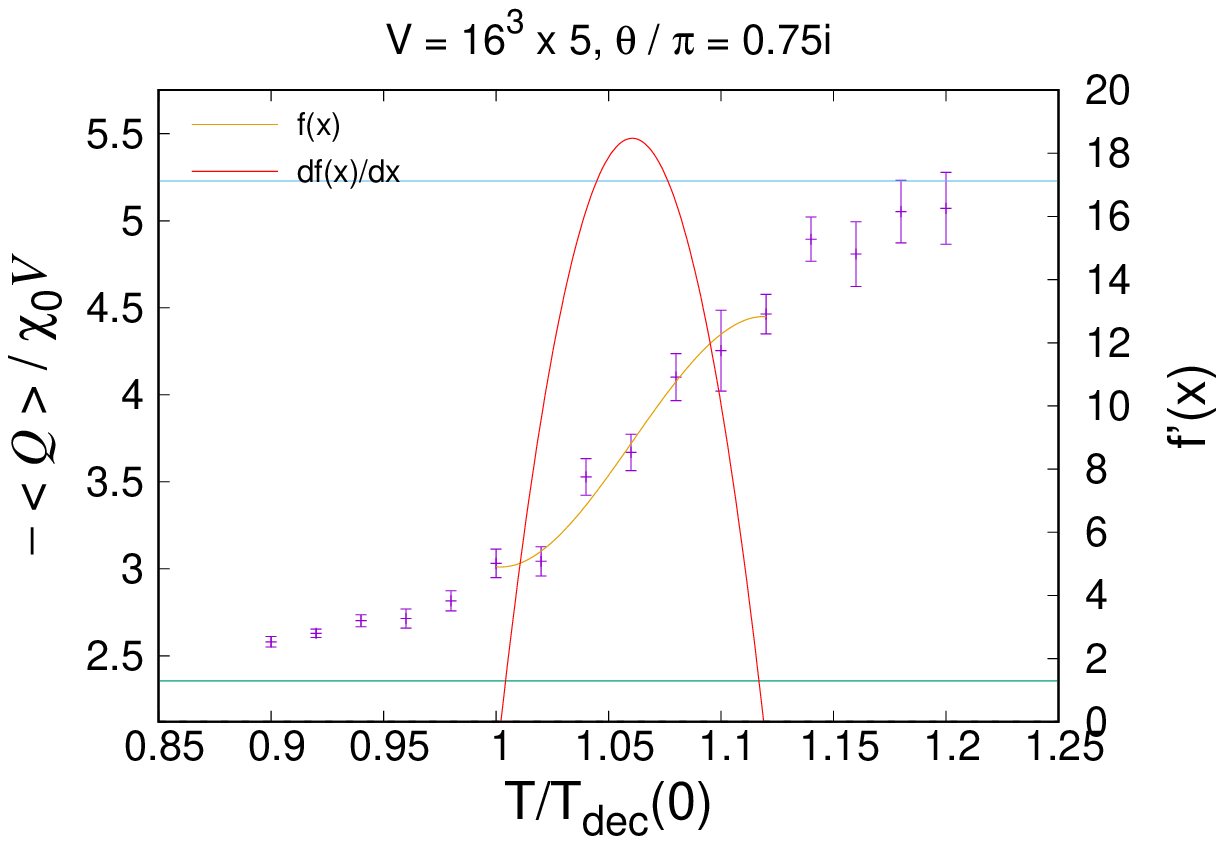}
\includegraphics[scale=0.55]{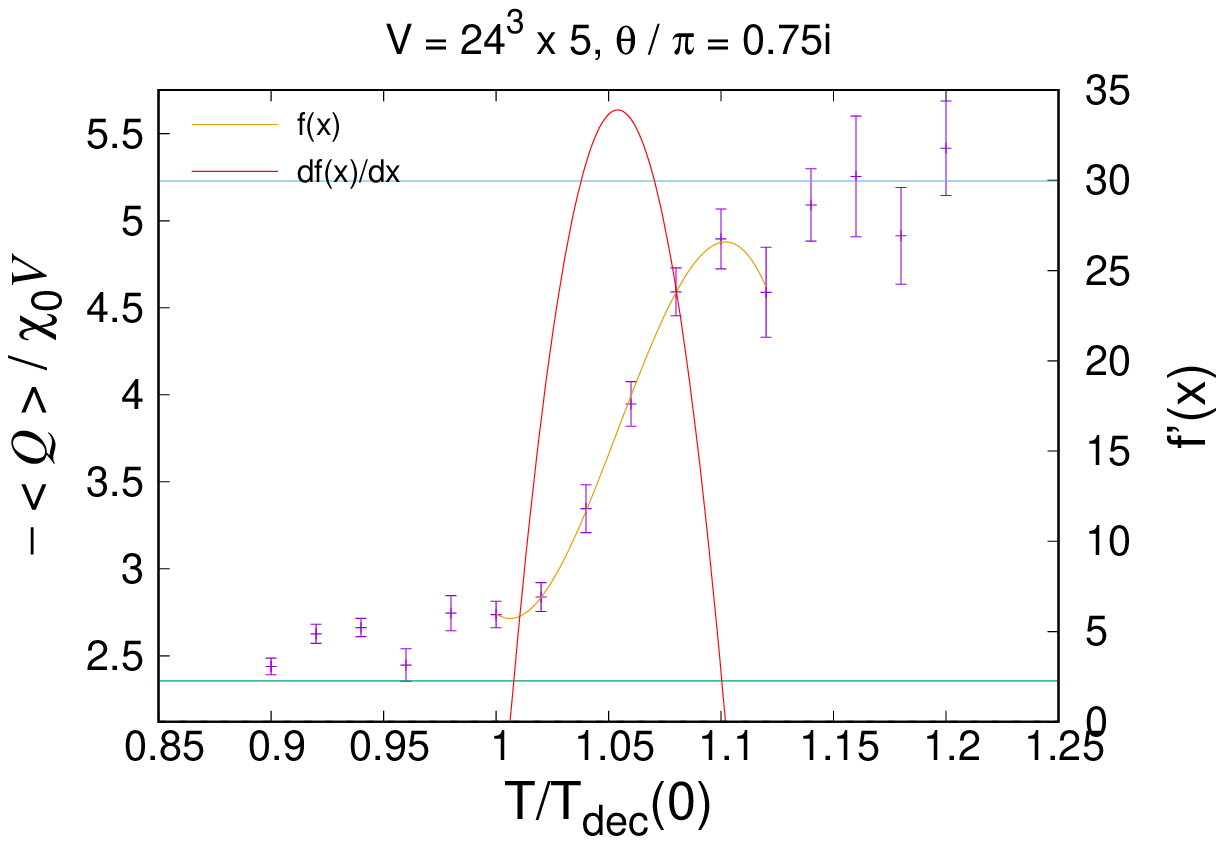}
\par\end{centering}
\caption{\label{fig:QT_var_Ls_0.75} The temperature dependence of $-\left\langle Q\right\rangle _{\tilde{\theta}}/\chi_{0}V$
at $\theta/\pi=0.75i$ for $V=16^{3}\times5$ (left) and $V=20^{3}\times5$
(right). We fit the data points by a cubic function $f(x)=ax^{3}+bx^{2}+cx+d$.
The orange solid line shows the result of fitting. The red solid line
is derivative of $f(x)$ with respect to $x$.}
\end{figure}

In order to see the temperature dependence of $-\langle Q\rangle_{\tilde{\theta}}/\chi_{0}V$,
we plot it against temperature at fixed $\theta/\pi=0.75i$ in Fig.~\ref{fig:QT_var_Ls_0.75}.
The left figure is the result for $V=16^{3}\times5$,
and the right figure is the result for $V=20^{3}\times5$. 
The yellow curve shows the result of fitting by a cubic function $f(x)=ax^{3}+bx^{2}+cx+d$
where $a,b,c$ and $d$ are fitting parameters.
The orange curve is the derivative of $f(x)$.
We find that the derivative 
is the largest at around $T_{{\rm peak}}\sim1.06T_{{\rm dec}}(0)$.
We also find that the height of the peak grows as the spatial volume
$V_{\mathrm{s}}$ increases. 

\begin{figure}
\begin{centering}
\includegraphics[scale=0.55]{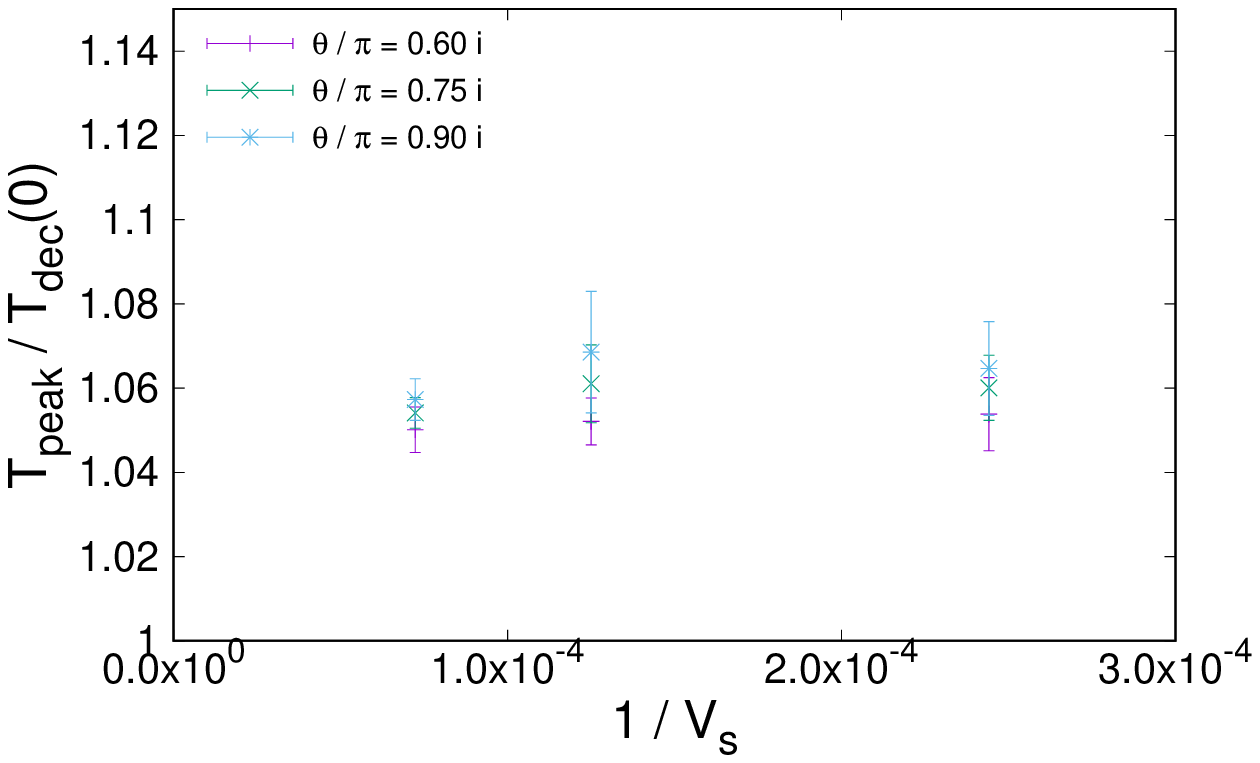}
\includegraphics[scale=0.55]{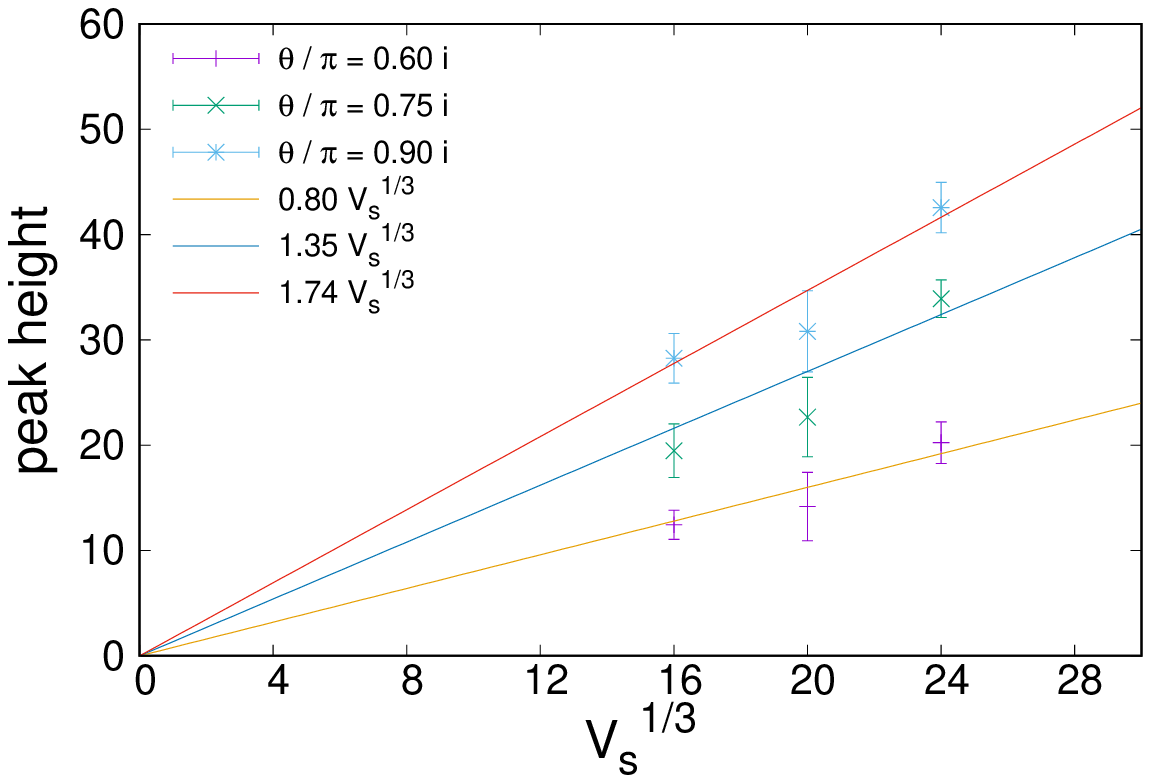}
\par\end{centering}
\caption{\label{fig:t_peak} (left) The peak position $T_{{\rm peak}}/T_{{\rm dec}}(0)$
against $1/V_{{\rm s}}$ for $\theta/\pi=0.6i$, $0.75i$ and $0.9i$, where
$V_{{\rm s}}$ is the spacial lattice volume. (right) The peak height
of $f^{\prime}(x)$ is plotted against $V_{{\rm s}}^{1/3}$. 
The straight lines represent the fit to the behavior $aV_{{\rm s}}^{1/3}$.}
\end{figure}

In Fig.~\ref{fig:t_peak}(left), we plot the peak position $T_{{\rm peak}}/T_{{\rm dec}}(0)$
against $1/V_{{\rm s}}$ obtained by the same analysis for $\theta/\pi=0.6i$,
$0.75i$ and $0.9i$. The significant volume dependence of $T_{{\rm peak}}$ is not observed. These
results suggest that there is a phase transition around $T/T_{{\rm dec}}(0)\sim1.06$.
In Fig.~\ref{fig:t_peak}(right), the peak height of the fitting
function is plotted against $V_{{\rm s}}^{1/3}$.
This non-linear finite size
scaling suggests that the phase transition is of the second order or
higher. 

The existence of the transition indicates that the distribution $\rho(q)$
of the topological charge changes drastically around $T/T_{{\rm dec}}(0)\sim1.06$.
Assuming that CP symmetry at $\theta=\pi$ is spontaneously broken
at low temperature, the drastic change of $\rho(q)$ should correspond
to the critical behavior of $\left\langle Q\right\rangle _{\theta=\pi}$.
Thus, we identify the critical temperature $T_{{\rm peak}}$ as the
CP restoration temperature $T_{\mathrm{CP}}$, which suggests $T_{\mathrm{CP}}>T_{{\rm dec}}(0)$. 

\section{Summary}

Recent studies on the 't Hooft anomaly matching condition for the 4D SU($N$) gauge
theory have suggested that the phase structure at $\theta=\pi$ should
be nontrivial. For large $N$, it is known that the CP symmetry
at $\theta=\pi$ is spontaneously broken in the confined phase, while
it is restored in the deconfined phase. However, for small $N$, a qualitatively
different phase structure can be realized, as long as the anomaly
matching condition is satisfied. In this work, we investigated this
issue for $N=2$ by hybrid Monte Carlo simulation of lattice
gauge theory. We probed the restoration of the CP symmetry by a
sudden change of the topological charge distribution at $\theta=0$,
which can be seen by simulating the theory with imaginary $\theta$.
This method is free from the sign problem. We measured the normalized expectation value $\left\langle Q\right\rangle /\chi_{0}V$
of the topological charge as a probe of the distribution. We found that
this observable has a finite-temperature transition around $T/T_{{\rm dec}}(0)\sim1.06$.

Although the deconfinement temperature $T_{{\rm dec}}$ at $\theta=\pi$
is not known, it is expected to be lower than $T_{{\rm dec}}(0)$.
Thus, our results suggest that the CP symmetry at $\theta=\pi$ is
restored at the temperature higher than the deconfinement temperature---unlike
the situation at large $N$. We plan to refine this result
by taking the continuum limit. We are also trying to extend this method
to the 4D SU(3) YM theory, in order to see a possible
qualitative difference between $N=2$ and $N=3$,
as suggested from the result in super YM theory.

\acknowledgments
The computations were carried out on the PC clusters in KEK Computing
Research Center and KEK Theory Center. This work is supported by the
Particle, Nuclear and Astro Physics Simulation Program No.2021-005
(FY2021) and No.2022-004 (FY2022) of Institute of Particle and Nuclear
Studies, High Energy Accelerator Research Organization (KEK).
A.~M. is supported by JSPS Grant-in-Aid
for Transformative Research Areas (A) JP21H05190.
M.~Honda. is supported by MEXT Q-LEAP, JST PRESTO Grant
Number JPMJPR2117 and JSPS Grant-in-Aid
for Transformative Research Areas (A) JP21H05190.
A.~Y. is supported by a JSPS Grant-in-Aid
for Transformative Research Areas (A) JP21H05191.

\bibliographystyle{JHEP}
\bibliography{ref}

\providecommand{\href}[2]{#2}\begingroup\raggedright\begin{thebibliography}{1}

\bibitem{Gaiotto:2017yup}
D.~Gaiotto, A.~Kapustin, Z.~Komargodski and N.~Seiberg, \emph{Theta, time
  reversal, and temperature},
  \href{https://doi.org/10.1007/JHEP05(2017)091}{\emph{JHEP} {\bfseries 05}
  (2017) 091} [\href{https://arxiv.org/abs/1703.00501}{{\ttfamily
  1703.00501}}].

\bibitem{Kitano:2017jng}
R.~Kitano, T.~Suyama and N.~Yamada, \emph{$\theta=\pi$ in $su(n)/\mathbb{Z}\_n$
  gauge theories}, \href{https://doi.org/10.1007/JHEP09(2017)137}{\emph{JHEP}
  {\bfseries 09} (2017) 137}
  [\href{https://arxiv.org/abs/1709.04225}{{\ttfamily 1709.04225}}].

\bibitem{Kitano:2020mfk}
R.~Kitano, N.~Yamada and M.~Yamazaki, \emph{{Is $N = 2$ Large?}},
  \href{https://doi.org/10.1007/JHEP02(2021)073}{\emph{JHEP} {\bfseries 02}
  (2021) 073} [\href{https://arxiv.org/abs/2010.08810}{{\ttfamily
  2010.08810}}].

\bibitem{Kitano:2021jho}
R.~Kitano, R.~Matsudo, N.~Yamada and M.~Yamazaki, \emph{{Peeking into the
  \ensuremath{\theta} vacuum}},
  \href{https://doi.org/10.1016/j.physletb.2021.136657}{\emph{Phys. Lett. B}
  {\bfseries 822} (2021) 136657}
  [\href{https://arxiv.org/abs/2102.08784}{{\ttfamily 2102.08784}}].

\bibitem{Chen:2020syd}
S.~Chen, K.~Fukushima, H.~Nishimura and Y.~Tanizaki, \emph{{Deconfinement and
  $\mathcal {CP}$ breaking at $\theta=\pi$ in Yang-Mills theories and a novel
  phase for SU(2)}},
  \href{https://doi.org/10.1103/PhysRevD.102.034020}{\emph{Phys. Rev. D}
  {\bfseries 102} (2020) 034020}
  [\href{https://arxiv.org/abs/2006.01487}{{\ttfamily 2006.01487}}].

\bibitem{DiVecchia:1981aev}
P.~Di~Vecchia, K.~Fabricius, G.~Rossi and G.~Veneziano, \emph{{Preliminary
  Evidence for U(1)-A Breaking in QCD from Lattice Calculations}}, .

\bibitem{Morningstar:2003gk}
C.~Morningstar and M.J.~Peardon, \emph{{Analytic smearing of SU(3) link
  variables in lattice QCD}},
  \href{https://doi.org/10.1103/PhysRevD.69.054501}{\emph{Phys. Rev. D}
  {\bfseries 69} (2004) 054501}
  [\href{https://arxiv.org/abs/hep-lat/0311018}{{\ttfamily hep-lat/0311018}}].

\end{thebibliography}\endgroup

\end{document}